# Ideal Graphene/Silicon Schottky Junction Diodes

Dhiraj Sinha and Ji Ung Lee[*]

*College of Nanoscale Science and Engineering, the State University of New York, Albany, NY-12203*

Abstract

The proper understanding of semiconductor devices begins at the metal-semiconductor interface. The metal/semiconductor interface itself can also be an important device, as Schottky junctions often forms when the doping in the semiconductors is low. Here, we extend the analysis of metal-silicon Schottky junctions by using graphene, an atomically thin semimetal. We show that a fundamentally new transport model is needed to describe the graphene-silicon Schottky junction. While the current-voltage behavior follows the celebrated ideal diode behavior, the details of the diode characteristics is best characterized by the Landauer transport formalism, suggesting that the injection rate from graphene ultimately determines the transport properties of this new Schottky junction.

[*]Author to whom correspondence should be addressed. Electronic mail: jlee1@albany.edu

Graphene is one amongst a class of two dimensional (2D) layered materials which has been intensively researched for possible post-CMOS device and other novel applications[1-4]. Graphene-semiconductor Schottky junction is one such system being studied for applications in photo detection[1, 5, 6], communications[7], solar cells[8, 9], barristers[10] and chemical and biological sensors[11]. Graphene Schottky junctions with several different semiconducting materials have been reported[12-14]. In order to design such devices with a wide range of applicability, it is critical to establish fundamental understanding of Schottky junctions formed using an atomically thin material. Here, we provide a new transport mechanism to describe the electrical properties of graphene-silicon (Si) Schottky junctions.

Graphene Schottky diodes studied in the literature have focused on electrical & optical characteristics, such as power conversion efficiency, photocurrent responsivity, spectral bandwidth, detection limits and speed[1, 7, 9, 15-17]. Despite these breakthroughs a basic graphene Schottky device with an ideal current-voltage characteristic is still uncommon, and the fundamental transport mechanism is poorly understood. Here, we fabricate ideal graphene/*n*-type silicon (*n*-Si) Schottky junction diodes and show a new transport mechanism to describe the ideal diode behavior. We find that traditional bulk approaches to analyzing our graphene-based Schottky diodes are inadequate. Instead, we find excellent agreement using the Landauer transport formalism that includes the finite density of states of graphene

The ideality factor ($\eta$) of any junction measures the degree to which defects mediate transport. *$\eta=1$* corresponds to an ideal Schottky diode with purely thermionic charge transport across the barrier. The $\eta$ values for graphene Schottky junctions reported to date are far from ideal and are in the range ~1.3-30[8, 12, 18]. A gate controlled Schottky barrier device was reported with an ideality of 1.1. Without any gate control, however, the rectification was limited to two only orders in magnitude due to high leakage currents[10]. A defect free interface between graphene and semiconductor is necessary for realizing an ideal rectification behavior[19, 20]. Some recent studies have identified the existence of inhomogeneity at the graphene and semiconductor junction[14] as one reason for the non-ideality. This can give rise to high leakage currents and poor values for $\eta$[18, 21]. Such impurities at graphene semiconductor interface could be remnants of metal contaminants from the copper (Cu) substrate used in Low-Pressure Chemical Vapor Deposition (LPCVD) technique to grow the graphene. The LPCVD process involves the use of cold rolled Cu foil as the substrate. It has been found that impurities such as Si, Ca, Ru, Pt, Ce are present on the surface of Cu foil, before graphene growth, that will likely remain on the graphene even after etching the Cu post growth.[22]

In this study we report on graphene/$n$-Si Schottky junction diodes demonstrating nearly ideal diode behavior, with $\eta$ ~1.08. Such diode behavior was achieved by minimizing the metallic impurities in graphene by etching the top ~700nm of Cu using ammonium persulfate solution (APS) for 90 seconds. Graphene growth is performed using the LPCVD technique[23]. Post growth, a thin layer of PMMA is coated on top of graphene. The Cu foil below graphene is then etched leaving the graphene supported on the PMMA membrane. To fabricate Schottky junctions using graphene, we prepared wafers with 20nm $SiO_2$ on $n$-Si wafer with an array of trenches cut into the oxide. Immediately before the graphene transfer, the silicon surface in the trench was cleaned with buffered oxide etch (BOE) etch solution with $NH_4F$:HF (6:1) ratio to remove any surface oxide. Graphene was then transferred onto this substrate using a dry transfer procedure reported by Petrone et al.[24]. Photolithography was used to pattern graphene across the trench and over the field oxide. The active device area was defined as the area of graphene making direct contact with the silicon in the trench. The graphene lying outside the trench on the oxide was subsequently contacted with 7/70 nm thick Cr/Au, and the back contact to silicon was formed using 100nm thick Al. The schematic of a typical device structure is shown in the inset of Fig. 1a. The devices fabricated in this manner are labeled type-SB01. In addition, we fabricated two other types of Schottky devices using the same $n$-Si wafers. In graphene/$n$-Si Schottky devices of type SB02, we used the identical fabrication process except that the initial etching of Cu was not performed. This is the typical method used in other studies[6, 17]. Finally, for our third set of devices, we fabricated bulk Schottky diodes using Cr/$n$-Si junctions. We fabricated over 40 identical devices for each type, and the data we present are typical of the results we observe.

First, we analyze the current-voltage ($J-V$) characteristics of these devices. The measured current density $J$ was normalized with respect to reverse saturation current ($J_O$), and plotted as $J/J_O - V$ curves (shown in Fig. 1a) for several type-SB01 and type-SB02 devices. Plotting the curves in this way, we clearly observe differences between type-SB01 and type SB02 devices: Type-SB01 devices exhibit lower ideality factor and a pronounced increases in the leakage current with reverse bias, a feature we attribute to lower interface defect states in our analysis below. The unnormalized $J-V$ curves are shown in Fig. 1b for each device type.

To better understand the three device types, we study the $J-V$ curves using the Crowell – Sze model for thermionic emission and diffusion of carriers over a barrier[25]. Accordingly, the $J-V$ characteristics of Schottky devices can be approximated by the following diode equation:

$$J = J_O \left(e^{\frac{qV}{\eta k_B T}} - 1\right); J_O = A^* T^2 e^{-\frac{\Phi_B}{k_B T}} \qquad (1).$$

$J_O$ is the reverse saturation current, $\eta$ is the diode ideality factor, $q$ is the elementary charge, $k_B$ is the Boltzmann's constant and $T$ is the temperature. A* is the effective Richardson's constant and $\Phi_B$ is the Schottky barrier height, defined in Fig. 3. Using Eq. 1, we determined $\eta$ by fitting the forward bias current of each device type (we also ensure that both forward and reverse bias characteristics are well represented by (1), particularly at low reverse bias). Fig. 1b shows the fits to the representative $J-V$ curves from all three device types. We note that the type-SB01 device has the lowest leakage current and nearly ideal behavior. The extremely low leakage current, which corresponds to an usually low A* value compared to that of the Cr/$n$-Si device, calls for an explanation. Here, we show the low value for A* arises from the finite Density of States (DOS) of graphene, which can be accounted for using the Landauer transport model. We are able to establish an excellent fit over a significant region in the $J-V$ curves at low voltage range. At higher forward bias, however, a significant deviation results from a series contact resistance, which is not included in our model. A relatively low ideality factor of $\eta$ ~1.08 characterizes type-SB01 devices, indicative of nearly defect-free interface. In contrast, a higher ideality value of ~1.5 characterizes the type-SB02 devices, which we attribute to the impurities from the Cu foil.

By varying the temperature, we extract the barrier height $\Phi_B$ using Eq. (1). The temperature dependent $J-V$ behavior was characterized between T=300-380K. We also determine the Richardson constant $A^*$ from the activation energy extracted through $J_O$ over the same temperature range. Fig. 2(a) shows the extracted barrier height as a function of reverse bias $V$ for all three device types. The barrier height for the Cr/*n*-Si junction is ~0.37 eV. This barrier height is largely constant with reverse bias, indicating that Fermi energy level in the semiconductor is pinned. A small modulation seen in Schottky barrier can be accounted for by the image charge barrier lowering[26]. The type-SB02 device showed a similar behavior with a largely constant barrier height of ~0.57 eV (excluding the small contribution from image charge barrier lowering). As discussed earlier, we associate the relatively constant barrier height with reverse bias in SB02 devices to pinning of the Fermi level due to impurities from Cu. This is consistent with the general properties of metal-semiconductor Schottky junctions with large interface states that pin the Fermi level, making $\Phi_B$ largely independent of bias[27, 28]. For the type-SB01 devices, however, we observe a noticeable lowering of the Schottky barrier height with increasing reverse bias. We discussion this below in the context of the origin of the ideal I-V behavior.

The A* extracted using the temperature dependent $J-V$ plots is strikingly different across these three devices (see Fig. 1b). In principle, the Richardson constant depends only on the material properties of Si, although it is known that the properties of the metal can cause some variations in its value[29]. For example, variations in A* have been reported in relation to inhomogeneity in barrier height, interfacial layer, quantum mechanical reflection and tunneling of carriers[30-32]. The A* for our Cr/*n*-Si junction is comparable to the values that have been reported by others[33, 34]. In contrast, in both graphene/*n*-Si junctions (type-SB01 and type-SB02), the A* values are significantly lower than the theoretical value of 112 ($A/cm^2/K^2$) for *n*-Si and much lower than those reported for most metals. Below, we provide a new analysis using the Landauer transport formalism to explain the low values of A*.

We first analyze the origin of A* and compared it to type SB01 devices, since the new transport formalisms leads to an ideal diode behavior. Next, we discuss the bias dependence on $\Phi_B$ for these devices, which we have not accounted for in our analysis so far. The new analysis provides an alternative to the ideal diode behavior used in conventional metal-semiconductor Schottky junctions. Our analysis suggests that the finite density of states of graphene plays a crucial role in determining the prefactor A*. The Landauer transport formalism is given as[35]

$$J = \frac{q}{\tau}\int_{-\infty}^{+\infty} T(E)D(E)(f_g - f_{Si})dE \quad (2),$$

where $q$ is the elementary charge, $\tau$ is the timescale for carrier injection from the contact, $T(E)$ is the transmission probability over the barrier energy $\Phi_B$, and $D(E) = \frac{2}{\pi(\hbar v_F)^2}|E| = D_0|E|$ is the density of states of graphene ( $v_F$ is the Fermi velocity in graphene and $\hbar$ is the reduced Planck's constant). $f_g$ and $f_{Si}$ are the Fermi-Dirac functions for graphene and silicon, respectively (shown in Fig. 3). Eq. 2 can be solved for both forward and reverse biases, provided that $\Phi_B \gg k_B T$ and $T(E)$ is defined as in Fig. 3. Here, we assume $T(E) = 1$ as long as the carrier energy is above $\Phi_B$. Eq. (2) then yields the ideal diode equation

$$J = J_o(e^{\frac{qV}{\eta k_B T}} - 1) \quad (3)$$

$$J_o = \left[\frac{q \cdot D_O}{\tau}(k_B T)^2 \left(\frac{\Phi_B}{k_B T} + 1\right)\right] e^{-\frac{\Phi_B}{k_B T}}$$

One crucial difference between $J_o$ in Eq. (3) to the one in Eq. (1) is the additional temperature difference we find in the Landauer formalism (in Eq. (1) A* is independent of T). To confirm our model, we fit the measured reverse-bias current for type-SB01 device using the Landauer transport model for $J_o$ in Eq. 3. Fig. 2(b) shows a comparison between Eqs. (1) and (3) to the measured $J_O$ $vs$ $T$ of type- SB01 device at three different reverse bias conditions. The goodness of fit established with the experimental data using each model is ~0.99. Fig. 2(b) demonstrates that the Richardson-Schottky model and the Landauer model are indistinguishable. This is because the dominant temperature dependence comes from $e^{-\frac{\Phi_B}{k_B T}}$. The crucial difference, however, is that our model can account for the low values for A* with $\tau = 4.62 \times 10^{-11}$ sec obtained from the fit. $\tau^{-1}$ represents the injection rate of carriers from the contact to graphene and is related to the coupling energy. A lower bound for $\tau$ can be determined from metals that make a very low contact resistance (larger coupling) to graphene. We extract a lower bound of $\tau = 1.3 \times 10^{-13}$ sec for Pd [36], which is consistent with the very low contact resistance that they measure. Here, our contacts are characterized by a much larger $\tau$ (smaller coupling energy), consistent with a large contact resistance of our devices. From Fig. 1b we estimate our contact resistance to be 1.18 MΩ-μm, which is considerably larger than the reported value of ~500 Ω-μm using Pd [36].

Finally we explain the bias dependence of barrier height $\Phi_B$ for the type SB01 device, shown in Fig. 2a, using the band diagram in Fig. 3. Our model is consistent with the ideal diode behavior of SB01 devices that is largely void of interface states that can pin the Fermi level. Since they exhibit low pinning, we anticipate that the Fermi level can be modulated by gating as the density of states in graphene is low near the Dirac point. We assume that the equilibrium Fermi level in graphene is at Dirac point ($E_{f_g}^O$) (Fig. 3 (left)). $\Phi_B^O = E_C - E_{f_g}^O$ defines the equilibrium barrier height. Applying a reverse bias voltage $V$ causes electrostatic doping in graphene through the depletion capacitance in silicon ($C_{Si}$). As a result of this extrinsic doping, Fermi level shifts upward to $E_{f_g}{'}$ (Fig. 3 (center)), lowering $\Phi_B$. The dependence of $E_{f_g}{'}$ on reverse bias voltage $V$ is established using the following equation.

$$q \int_{-\infty}^{+\infty} D(E) f(E_f) dE = C_{Si} |V| \quad (4).$$

Here $C_{Si}$, the depletion capacitance given by $dQ/dV$, is the incremental change in charge per unit area due to change in applied voltage across the depletion layer in Si. In this case, using a $10^{16}/cm^3$ doping for $n$-Si and an experimentally measured barrier height of $\Phi_B = 0.62$ eV, the built-in voltage due to space charge in the depletion layer works out to be 0.42 eV ($V_{bi} = \Phi_B^O - (E_C - E_{f_s})$). The amount of charge corresponding to the $V_{bi}$ is then used to estimate $C_{Si}$ – which is approximated to be $5x10^{-8}$ F/cm² [6]. $f(E_f)$ is the Fermi function in graphene, and $V < 0$ is the reverse-bias voltage. Modulation of Fermi level in graphene relative to the Dirac point is manifested as a reduction in effective barrier height by $\Delta\Phi_B(V) = |E_{f_g}^o - E'_{f_g}(V)|$. The reduced barrier height as a function of reverse-bias voltage is specified as $\Phi_B'(V) = \Phi_B^O - \Delta\Phi_B(V)$. Inserting $\Phi_B'(V)$ in Eq. 3, we obtain a diode equation which incorporates the voltage-dependent diode characteristics observed in type-SB01 devices. Fig. 4 shows a comparison between data and our model that now includes the doping induced barrier lowering.

In summary, we demonstrate ideal graphene/*n*-Si Schottky diodes by growing graphene on etched Cu foils. This is believed to significantly reduce impurities from the Cu foil that cause pinning of the Fermi level and reduce trap states that result in ideal diodes. The lack of pinning is demonstrated in the strong bias dependence on the leakage current, which we can capture using a simple capacitance model. Finally, we provide a new Schottky junction model based on the Landauer transport formalism that explains the unusually low value of the Richardson constant.

# References

1. An, X.; Liu, F.; Jung, Y. J.; Kar, S. *Nano letters* **2013,** 13, (3), 909-916.
2. Sung, C.-Y.; Lee, J. U. *Spectrum, IEEE* **2012,** 49, (2), 32-59.
3. Britnell, L.; Gorbachev, R.; Jalil, R.; Belle, B.; Schedin, F.; Mishchenko, A.; Georgiou, T.; Katsnelson, M.; Eaves, L.; Morozov, S. *Science* **2012,** 335, (6071), 947-950.
4. Mueller, T.; Xia, F.; Avouris, P. *Nature Photonics* **2010,** 4, (5), 297-301.
5. Zeng, L.-H.; Wang, M.-Z.; Hu, H.; Nie, B.; Yu, Y.-Q.; Wu, C.-Y.; Wang, L.; Hu, J.-G.; Xie, C.; Liang, F.-X. *ACS applied materials & interfaces* **2013,** 5, (19), 9362-9366.
6. An, Y.; Behnam, A.; Pop, E.; Ural, A. *Applied Physics Letters* **2013,** 102, (1), 013110.
7. Lv, P.; Zhang, X.; Deng, W.; Jie, J. *Electron Device Letters, IEEE* **2013,** 34, (10), 1337 - 1339
8. Miao, X.; Tongay, S.; Petterson, M. K.; Berke, K.; Rinzler, A. G.; Appleton, B. R.; Hebard, A. F. *Nano letters* **2012,** 12, (6), 2745-2750.
9. Li, X.; Zhu, H.; Wang, K.; Cao, A.; Wei, J.; Li, C.; Jia, Y.; Li, Z.; Li, X.; Wu, D. *Advanced Materials* **2010,** 22, (25), 2743-2748.
10. Yang, H.; Heo, J.; Park, S.; Song, H. J.; Seo, D. H.; Byun, K.-E.; Kim, P.; Yoo, I.; Chung, H.-J.; Kim, K. *Science* **2012,** 336, (6085), 1140-1143.
11. Kim, H.-Y.; Lee, K.; McEvoy, N.; Yim, C.; Duesberg, G. S. *Nano letters* **2013,** 13, (5), 2182-2188.
12. Chen, C.-C.; Aykol, M.; Chang, C.-C.; Levi, A.; Cronin, S. B. *Nano letters* **2011,** 11, (5), 1863-1867.
13. Tongay, S.; Lemaitre, M.; Schumann, T.; Berke, K.; Appleton, B.; Gila, B.; Hebard, A. *Applied Physics Letters* **2011,** 99, (10), 102102.
14. Tongay, S.; Lemaitre, M.; Miao, X.; Gila, B.; Appleton, B.; Hebard, A. *Physical Review X* **2012,** 2, (1), 011002.
15. Singh, R. S.; Nalla, V.; Chen, W.; Wee, A. T. S.; Ji, W. *ACS nano* **2011,** 5, (7), 5969-5975.
16. Nie, B.; Hu, J. G.; Luo, L. B.; Xie, C.; Zeng, L. H.; Lv, P.; Li, F. Z.; Jie, J. S.; Feng, M.; Wu, C. Y. *Small* **2013,** 9, (17), 2872-2879.
17. Ural, A. In *Fabrication and Characterization of Photodetectors Composed of Graphene/Silicon Schottky Junctions*, 225th ECS Meeting (May 11-15, 2014), 2014; Ecs.
18. Yim, C.; McEvoy, N.; Duesberg, G. S. *Applied Physics Letters* **2013,** 103, (19), 193106.
19. Rajput, S.; Chen, M.; Liu, Y.; Li, Y.; Weinert, M.; Li, L. *Nature communications* **2013,** 4, 2752.
20. Shivaraman, S.; Herman, L. H.; Rana, F.; Park, J.; Spencer, M. G. *Applied Physics Letters* **2012,** 100, (18), 183112.
21. Tongay, S.; Schumann, T.; Hebard, A. *Applied Physics Letters* **2009,** 95, (22), 222103.
22. Kim, S. M.; Hsu, A.; Lee, Y.-H.; Dresselhaus, M.; Palacios, T.; Kim, K. K.; Kong, J. *Nanotechnology* **2013,** 24, (36), 365602.
23. Li, X.; Cai, W.; An, J.; Kim, S.; Nah, J.; Yang, D.; Piner, R.; Velamakanni, A.; Jung, I.; Tutuc, E. *Science* **2009,** 324, (5932), 1312-1314.
24. Petrone, N.; Dean, C. R.; Meric, I.; Van Der Zande, A. M.; Huang, P. Y.; Wang, L.; Muller, D.; Shepard, K. L.; Hone, J. *Nano letters* **2012,** 12, (6), 2751-2756.
25. Crowell, C.; Sze, S. *Solid-state electronics* **1966,** 9, (11), 1035-1048.
26. Rhoderick, E. H.; Williams, R., *Metal-semiconductor contacts*. Clarendon Press Oxford: 1988.
27. Bardeen, J. *Physical Review* **1947,** 71, (10), 717.
28. Cowley, A.; Sze, S. *Journal of Applied Physics* **1965,** 36, (10), 3212-3220.
29. Modinos, A., Secondary Electron Emission Spectroscopy. In *Field, Thermionic, and Secondary Electron Emission Spectroscopy*, Springer: 1984; pp 327-345.
30. Horváth, Z. J. *Solid-State Electronics* **1996,** 39, (1), 176-178.
31. Dökme, İ.; Altindal, Ş.; Bülbül, M. M. *Applied surface science* **2006,** 252, (22), 7749-7754.

32. Taşçıoğlu, İ.; Aydemir, U.; Altındal, Ş. *Journal of Applied Physics* **2010,** 108, (6), 064506.
33. Martinez, A.; Esteve, D.; Guivarc'h, A.; Auvray, P.; Henoc, P.; Pelous, G. *Solid-State Electronics* **1980,** 23, (1), 55-64.
34. Anantha, N.; Doo, V.; Seto, D. *Journal of The Electrochemical Society* **1971,** 118, (1), 163-165.
35. Datta, S., *Quantum transport: atom to transistor*. Cambridge University Press: 2005.
36. Xia, F.; Perebeinos, V.; Lin, Y.-m.; Wu, Y.; Avouris, P. *Nature nanotechnology* **2011,** 6, (3), 179-184.

**Figures**

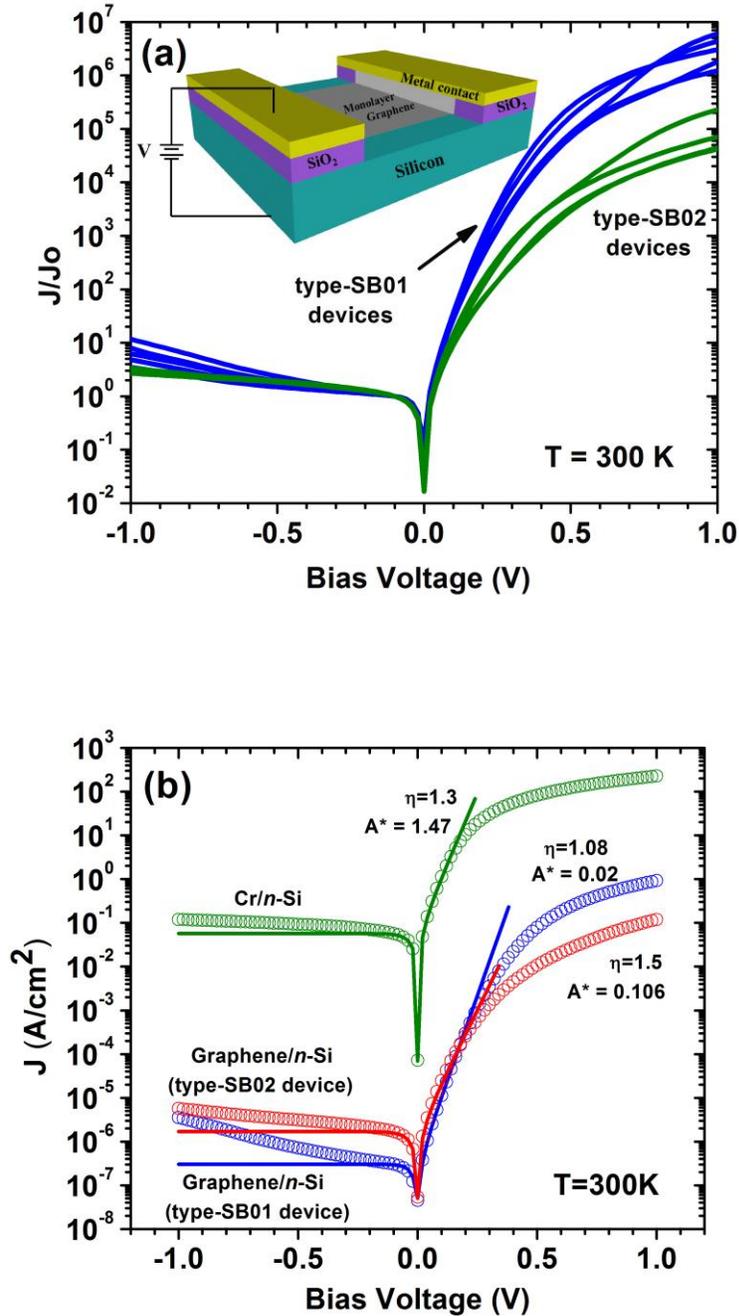

FIG.1: (a) Plot of the normalized current – voltage characteristics of graphene/$n$-Si Schottky diodes of type-SB01 and type-SB02. The inset shows a schematic diagram of the device. (b) Plot of the unnormalized current – voltage characteristics (open circles) and the fit (solid line) obtained using Schottky diode model. A* is the effective Richardson constant in $A/cm^2/K^2$.

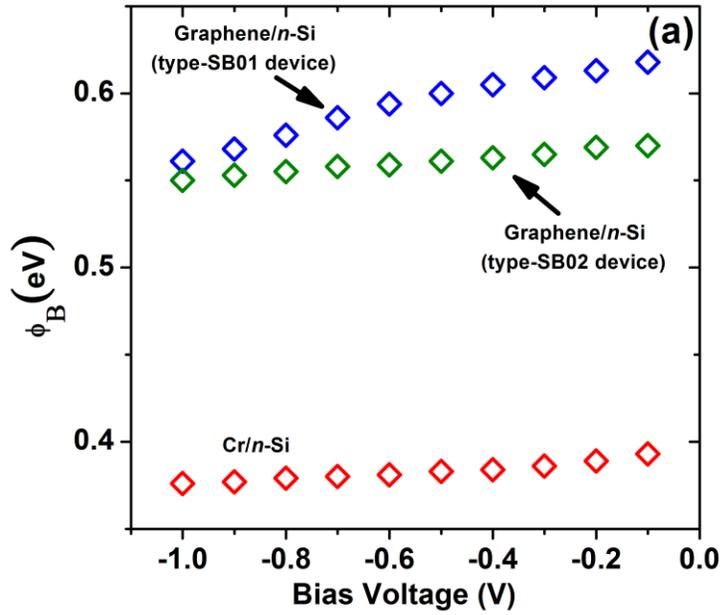

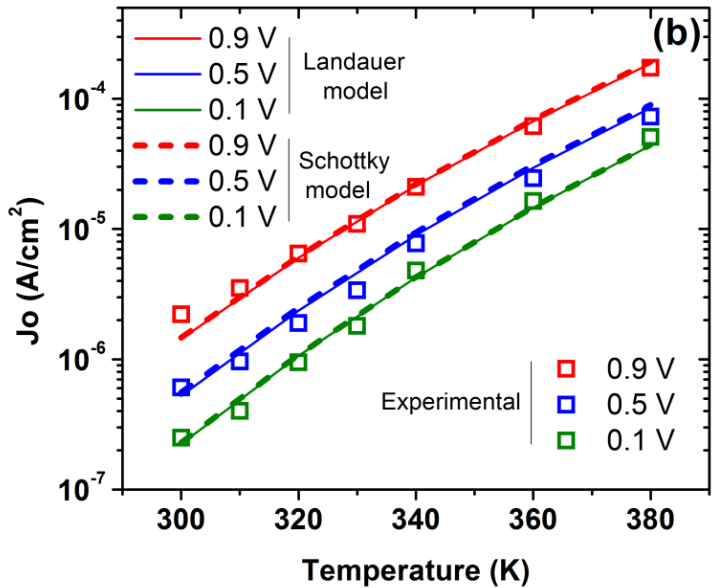

FIG. 2: (a) Plot of the extracted Schottky barrier height $\Phi_B$ as a function of reverse bias voltage $V$. (b) The temperature dependence of leakage current for type-SB01 device, obtained from experimental measurements (open squares), along with fits to the Schottky diode model (dashed lines) and Landauer model (solid lines).

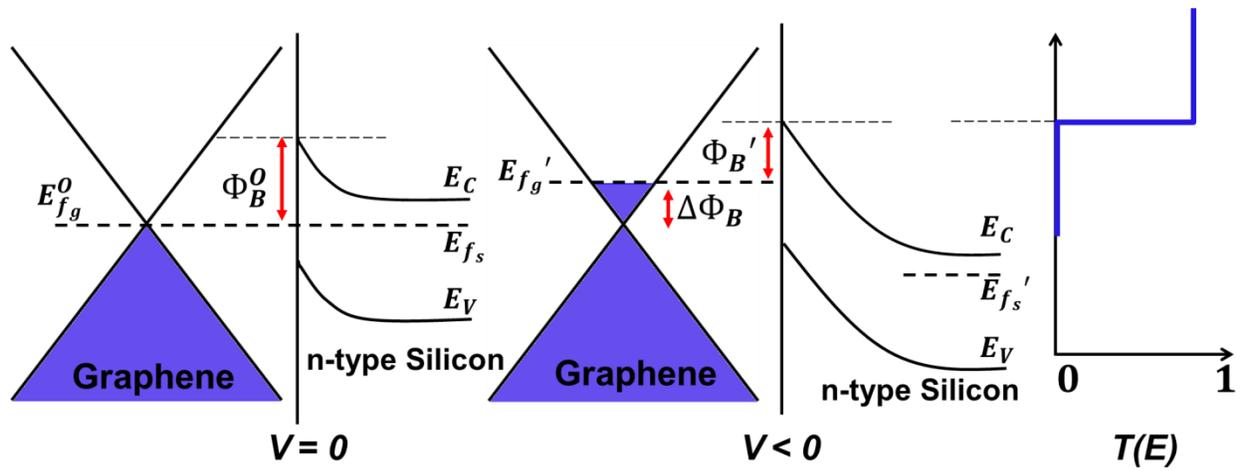

FIG. 3: (left) The energy band diagram of type-SB01 graphene/$n$-Si junction at equilibrium. $E_C$, $E_{f_s}$ and $E_V$ are the conduction band, Fermi level and valence band of Si, respectively. $E_{f_g}^0$ is the equilibrium Fermi level in graphene, and $\Phi_B^O = E_C - E_{f_g}^O$ is the barrier height. (center) $E'_{f_g}$ and $\bar{E}'_{f_s}$ show the Fermi levels under reverse bias. The upward shift of Fermi level in graphene causes lowering of the Schottky barrier $\Phi_B$, defined as $\Phi_B' = \Phi_B^O - \Delta\Phi_B$. (c) Illustrates the transmission probability of electrons over the Schottky barrier.

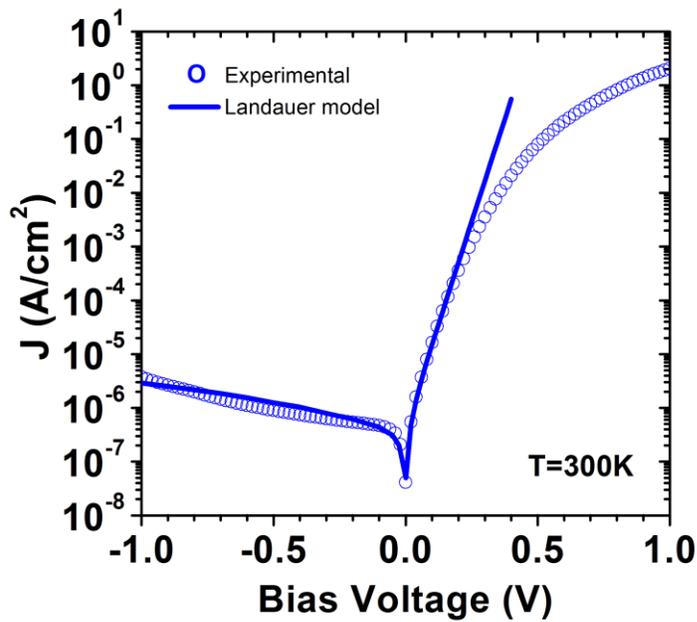

FIG. 4: Current – Voltage characteristics obtained using the Landauer transport model with barrier lowering (solid) showing excellent agreement with measured characteristics.

13